\documentclass[aps,prl,twocolumn,superscriptaddress,longbibliography]{revtex4-2}

\usepackage{graphicx}
\usepackage{dcolumn}
\usepackage{bm}
\usepackage{color}
\usepackage[colorlinks=true,allcolors=blue,breaklinks=true]{hyperref}
\usepackage{amsmath} 

\newcommand{\avg}[1]{\left\langle #1 \right\rangle}

\begin{document}

\title{Sheared Amorphous Packings Display Two Separate Particle Transport Mechanisms}


\author{Dong Wang}
\thanks{These authors contributed equally}
\affiliation{Department of Physics \& Center for Non-linear and Complex Systems, Duke University, Durham, North Carolina 27708, USA}

\author{Joshua A. Dijksman}
\thanks{These authors contributed equally}
\affiliation{Physical Chemistry and Soft Matter, Wageningen University \& Research, Stippeneng 4, 6708 WE Wageningen, The Netherlands}

\author{Jonathan Bar\'{e}s}
\affiliation{LMGC, UMR 5508 CNRS-University Montpellier, 34095 Montpellier, France}

\author{Jie Ren}
\affiliation{Merck \& Co., Inc., West Point, Pennsylvania 19486, USA}

\author{Hu Zheng}
\email{zhenghu@tongji.edu.cn}
\affiliation{Department of Physics \& Center for Non-linear and Complex Systems, Duke University, Durham, North Carolina 27708, USA}
\affiliation{Department of Geotechnical Engineering, College of Civil Engineering, Tongji University, Shanghai, 200092, China}

\date{\today}

\begin{abstract}
Shearing granular materials induces non-affine displacements. Such non-affine displacements have been studied extensively, and are known to correlate with plasticity and other mechanical features of amorphous packings. A well known example is shear transformation zones as captured by the local deviation from affine deformation, $D^2_{min}$, and their relevance to failure and stress fluctuations. We analyze sheared frictional athermal disk packings and show that there exists at least one additional mesoscopic transport mechanism that superimposes itself on top of local diffusive motion. We evidence this second transport mechanism in a homogeneous system via a diffusion tensor analysis and show that the trace of the diffusion tensor equals the classic $D^2_{min}$ when this second mesoscopic transport is corrected for. The new transport mechanism is consistently observed over a wide range of volume fractions and even for particles with different friction coefficients and is consistently observed also upon shear reversal, hinting at its relevance for memory effects. 
\end{abstract}

\maketitle

Sheared disordered materials have peculiar mechanical properties, including the tendency to either dilate or contract with strain \cite{reynolds85,bi11nature,ren13prl,pouliquen03prl,tsai03prl,kabla09prl,toiya04prl}, display peak stress behavior~\cite{oda1982experimental} and remember the way they were mechanically excited in their past \cite{2019rmp, memsand1999, 2005prehecke, peshkov19pre}. These mechanical nontrivialities are most pronounced when a packing of particles is almost but not entirely rigid, as set by the particle volume fraction. Around this volume fraction, shear has the ability to ``jam'' or rigidify disordered packings \cite{bi11nature, Zhao2019_prl}. The underpinnings of the mechanism are still under intense debate \cite{sarkar15pre,kumar16gm,bertrand17pre,baity-jessi17jstatphys,wang2018_prl,xiong19gm}. One of the complicating factors is that many of the nontrivial mechanical features are transient phenomena, making them particularly difficult to probe, as they are highly dependent on packing preparation, boundary conditions and other seemingly insignificant details of the system \cite{toiya04prl,kumar16gm}. In the past, qualitative and quantitative features such as shear transformation zones \cite{falk98pre,li15pre,lois05pre,lebouil14prl}, dynamical heterogeneities \cite{mehta2008pnas,berthier2011dynamical,bares2017_pre}, anisotropy \cite{chacko18jfm,misra16cmt,poorsolhjouy19jmps} and vortices \cite{conway2004nature} have all been observed in the transient response of amorphous packings exposed to shear. However, it is not clear which of these transport phenomena are relevant for the observed mechanics, or how many of these occur simultaneously during shear.

We show here that particle displacements during the startup transient in a sheared amorphous packing is dominated by two simultaneous but \emph{distinct} particle transport mechanisms, only one of which is sensitive to the particle volume fraction. Besides the locally non-affine, diffusive particle motion as quantified by the standard $D^2_{min}$ \cite{falk98pre,li15pre}, we observe a directional, collective particle transport mechanism that occurs on a length scale of up to ten times the shear transformation zone size. This directional transport, from now on referred to as $D^1$, occurs before the emergence of dilatancy-induced Reynolds pressure yet disappears in the random close packing limit. Moreover, the onset of $D^1$ emerges at larger $\phi$ for lower particle friction coefficients, which hints at its relevance for shear jamming phenomena, which display similar trends \cite{ren13prl, Zhao2019_prl}. Interestingly, $D^1$ survives cyclic shear over many different cycles, converges to limit cycle behavior and displays strain dependence in its response to cyclic driving, making $D^1$ of interest as metric to capture the dynamics of memory formation in amorphous packings \cite{gadala-maria80jr, toiya04prl, keim11prl, paulsen14prl, kumar16gm, fiocco14prl, royer15pnas, kou17nature, peshkov19pre}.


To observe particle transport in disordered media, we use a setup introduced before \cite{ren13prl,wang2018_prl}.  We apply simple shear to bidisperse granular systems composed of photoelastic discs with different inter-particle friction coefficients, $\mu$. Shear is performed in an apparatus that suppresses shear bands and related shear-induced density fluctuations entirely \cite{ren13prl}. The optical properties of photoelastic discs reveals particle-scale contact forces when placed between a pair of crossed polarizers \cite{howell97pg,howell99prl,zadeh19_gm}. Disks were cut from photoelastic sheets (Vishay PSM-4), resulting in a $\mu \approx 0.7$ \cite{ren13prl,zadeh19_gm}. Different $\mu$ were achieved by either wrapping these particles with Teflon tape, reducing $\mu$ to $\approx 0.15$, or making another set of photoelastic disks with fine teeth on the edge \cite{zadeh19_gm}. For a picture of the particles used, see Supplemental Material (SM) \cite{sm}. Henceforth, we will refer to particles with $\mu \approx 0.15$, $\mu \approx 0.7$, and fine teeth as $\mu = \mu_l$, $\mu = \mu_m$, and $\mu = \mu_h$ particles, respectively. The system contained approximately $1000$ discs with a large to small number ratio $1:3.3$ to prevent crystallization. Every run at given packing fraction $\phi$ was repeated five times; the initial stress-free state was prepared anew for each run. For every friction coefficient, we explored a range of $\phi$ in which we could observe the emergence of dilatancy pressure within the strain amplitude achievable by the apparatus.

\begin{figure}
\centering
\includegraphics[width=\linewidth]{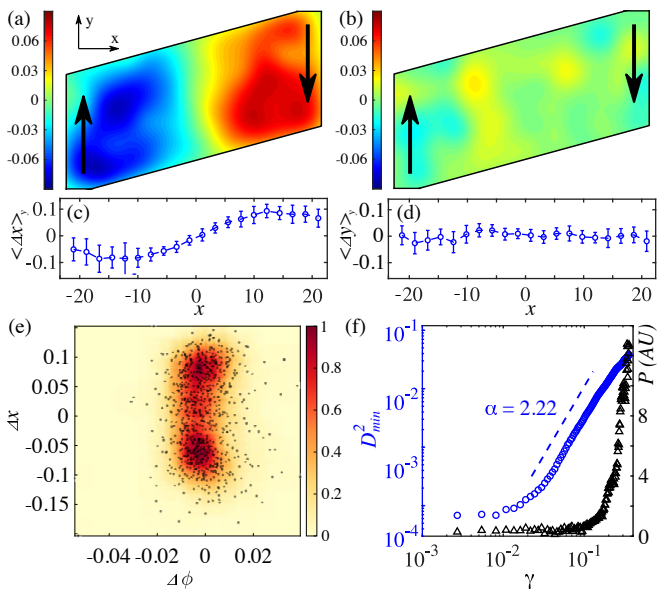}
\caption{Example of the system response to simple shear at packing fraction $0.78$. (a-b) Coarse grained non-affine $x$ and $y$ displacements after a shear strain $\gamma = 0.054$. The Cartesian coordinates are given in (a). The black arrows indicate the applied shear direction.(c-d) Non-affine $x$ and $y$ displacements averaged along the $y$ axis, as a function of $x$. Data are from the same states as (a) and (b), respectively. The dashed lines are a guide to the eye. $x = 0$ is at the center of the shear box. (e) Particles' non-affine $x$ displacements as a function of their local packing fraction change for each particle. The colored mask shows the normalized probability distribution of the points with the probability indicated by the color bar. (f) Local deviation from the affine deformation $D^2_{min}$ (left, blue circles) and system pressure (right, black triangles) as a function of shear strain $\gamma$.}
\label{fig-example}
\end{figure}

Shear was applied quasi-statically in the $y$ direction to a shear box (Fig.~\ref{fig-example}(a)). Starting from a stress-free state, the system was sheared by strain step of $\delta\gamma = 0.0027$. Then the system was left to relax for six seconds, followed by taking two images, which reveal information on particle position and photoelastic response. Such a process -- stepwise shearing, relaxing and imaging -- was repeated until a certain amount of total strain was achieved. The maximum $\gamma$ that can be achieved here is $0.54$, which is the limit of the experimental setup. From the images, we tracked particle positions and computed particle-scale and system-wide pressure by averaging light intensity gradient squared \cite{howell97pg,howell99prl,zadeh19_gm}. Henceforth, we use the average diameter of particles in the system $\bar{D}$ as the unit length in all results.


\begin{figure}
\centering
\includegraphics[width=\linewidth]{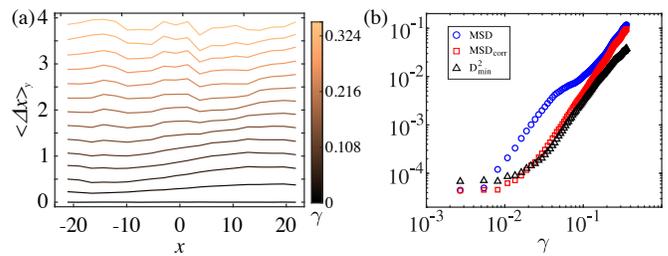}
\caption{(a) Non-affine $x$ displacements $\Delta x$, averaged in the same way as Fig.~\ref{fig-example}(c), as a function of the bin position for different shear strains $\gamma$. Different $\gamma$'s are indicated by the colorbar and the curves are shifted for better visibility. (b) Mean squared displacement (MSD) before (triangles) and after (circles) correcting the diffusion tensor $T$ with $\avg{\Delta x}_y$ and $\avg{\Delta y}_y$ from the non-affine displacements, and $D^2_{min}$ (squares), vs. $\gamma$. Data from the same run in Fig.~\ref{fig-example}.}
\label{fig-correction}
\end{figure}

Two transport mechanisms can be qualitatively observed under all experimental conditions. We can quantify the transport by decomposing the displacement field into the $x$ and $y$ component, which is referred to as the shear and perpendicular directions respectively (see Fig.~\ref{fig-example}(a)). We compute particle displacements at any given $\gamma$ compared to $\gamma = 0$. Fig.~\ref{fig-example}(a-b) show typical results of particle non-affine $x$ and $y$ displacements at $\phi=0.78$ and $\gamma = 0.054$, coarse-grained by a Gaussian function with a half width of $2\bar{D}$ \cite{zhang10ptps,clark2012gm}. In the $x$-direction, the expected affine displacement field is zero, but we see a substantial amount of particle displacement \emph{away} from the center of the box. For the $y$ direction, the expected global affine motion profile is linear; this have already been verified \cite{ren13prl}. Subtracting the globally affine linear motion reveals that on top of the linear deformation profile, a small perturbation is visible in Fig~\ref{fig-example}(b); however the perturbation is spatially more heterogeneous and much smaller than the transport in the $x$ direction, as evidenced by the mean transport $\avg{\Delta x}_y(x)$, $\avg{\Delta y}_y(x)$ across the system that are averaged over $y$ direction, as shown in Fig~\ref{fig-example}(c-d). Hence, from now on, we only look at the binned $x$-dependent displacement data, i.e., $\avg{\Delta x}_y(x)$ and $\avg{\Delta y}_y(x)$.

We can see that the displacements induced by the shear are not eased by local free volume availability. $\avg{\Delta x}_y$ is clearly a function of $x$, and one might relate this dependence with the local deviations $\Delta\phi$ from the mean density $\phi$. However, we see no correlation between $\Delta x$ and $\Delta\phi$ when we plot them together as shown in Fig.~\ref{fig-example}(e). Local non-affine displacements should be induced by local pressure fluctuations \cite{losert00prl}, yet as shown by Fig.~\ref{fig-example}(f) we find that the total non-affine motion induced by shear is not sensitive to the average global pressure in the system. In Fig.~\ref{fig-example}(f), we quantify the overall non-affine displacement with the standard $D^2_{min}$ metric and see that it rises independently of the evolving Reynolds pressure in the system. Here $D^2_{min}$ is calculated by averaging over all the particles the following quantity, as done previously \cite{falk98pre,li15pre}: 
\begin{equation}
    D^2_{min,i}(\gamma) = \frac{1}{N_i}\min_{\mathbf{J}_i}\sum_{j\in N_i}[\mathbf{r}_{ij}(\gamma)-\mathbf{J}_i\mathbf{r}_{ij}(\gamma=0)]^2,
\end{equation}
where $j$ is the index of all $N_i$ particles that are within $2\bar{D}$ distance to particle $i$ center, $\mathbf{r}_{ij}(\gamma)$ is the vector from particle $i$ to particle $j$ at a given strain $\gamma$, and $\mathbf{J}_i$ is the fitted strain field that minimizes the above quantity.

In Fig.~\ref{fig-correction}(b) and \ref{fig-example}(f), we observe that $D^2_{min} = \avg{D^2_{min,i}}_i$ dynamics in the packings is entirely consistent with earlier work on a completely different system~\cite{li15pre}, even though we use here a constant volume setting and observe an additional mesoscopic transport mechanism (see more details in SM \cite{sm}). We conclude that $D^2_{min}$ is a local background process that is largely insensitive to the details of the deformations imposed.

\emph{Identifying two different mechanisms ---} We extract the additional mesoscopic transport mechanisms that occur simultaneously with $D^2_{min}$ by measuring $\avg{\Delta x(x,\gamma)}_y$; see Fig.~\ref{fig-correction}(a). We observe that the directional displacement emerges immediately at small $\gamma$ but later becomes more spatially heterogeneous over smaller length scales (see details in SM \cite{sm}). To further show that the directional transport $D^1$ is independent of locally non-affine ``random'' motion, we compute the diffusion tensor of the particle displacement
\begin{equation}
    T = \langle \begin{pmatrix}
            \Delta x^2 & \Delta x \cdot \Delta y \\
            \Delta y \cdot \Delta x & \Delta y^2
            \end{pmatrix} \rangle,
\end{equation}
where $\avg{\cdots}$ is the average over all particles. We compute this tensor in two ways: once before and once after subtracting the local mesoscale displacement fluctuations $\avg{\Delta x}_y,\avg{\Delta y}_y$ from the displacement of the particles. From these two tensors we compute the trace and call them $\mathit{MSD}$ and $\mathit{MSD_{corr}}$ respectively. We note that subtracting only $D^1$ and not $\avg{\Delta y}_y$ gives similar results and the symmetric off-diagonal term is non-zero, which indicates that the major diffusion directions are not along $x$ or $y$ axes (see SM \cite{sm}). The result in Fig.~\ref{fig-correction}b indicates that after subtracting the mesoscale displacements, the trace of the diffusion tensor coincides with the $D^2_{min}$ metric: without mesoscale displacements, non-affine motion cannot be distinguished from diffusive behavior, consistent with earlier results~\cite{li15pre}.

\begin{figure}[t]
\centering
\includegraphics[width=\linewidth]{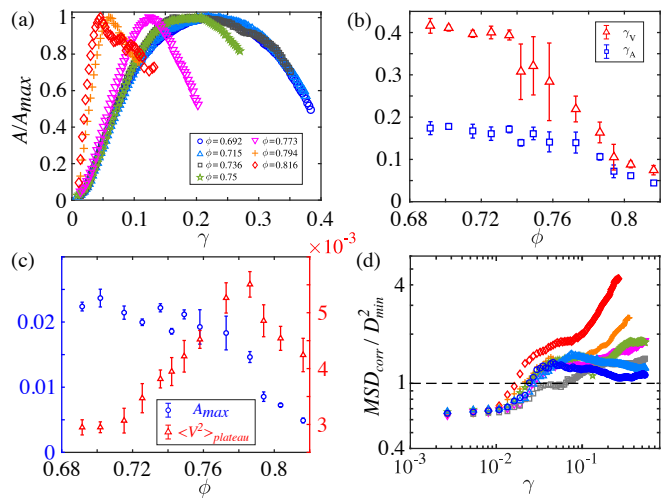}
\caption{(a) Linear slope $A$ extrapolated from $\avg{\Delta x}_y$ vs. $x$ for $x$ in the range $(-10, 10)$, normalized by its maximum value in a run $A_{max}$, vs. shear strain $\gamma$, at different $\phi$ as indicated by different symbols and colors. (b) Corresponding $\gamma$ at $A_{max}$ (left, circles), and $\gamma_{onset}$ for average vorticity $\avg{V^2}$ to reach the plateau (right, triangles), and (c) $A_{max}$ (left, circles), and plateau value of $\avg{V^2}$ (right, triangles), vs. $\phi$, vs. $\phi$, averaged over $5$ runs for each $\phi$. (d) The ratio between $\mathit{MSD_{corr}}$ and $D^2_{min}$ vs. $\gamma$, at various packing fractions $\phi$ as indicated the same way in (a). The dashed line corresponds to the ratio being 1.}
\label{fig-phi}
\end{figure}

One expects the strain needed to induce particle transport to decrease with $\phi$; this is indeed the case. Figure~\ref{fig-phi}(a) shows the amplitude of $D^1$ as measured by $A = d\avg{\Delta x}_y/dx |(x=0)$ as a function of strain, measuring the spatial gradient of the orthogonal displacement field in the center of the shear box. $A$ peaks at $A_{max}$, and this peak is reached at progressively smaller strain amplitudes (denoted as $\gamma_A$) as $\phi$ increases, as can be seen in Fig.~\ref{fig-phi}(b). However, the amplitude of $D^1$ is largest at small $\phi$. As shown in Fig.~\ref{fig-phi}(c), $A_{max}$ drops with increasing $\phi$ and seems to disappear around the random close packing limit of $\sim 0.83$, where also the strain needed to reach peak $D^1$ vanishes. Our data indicates the $D^1$ only disappears at packing fractions at which it is no longer possible to create stress free initial configurations. This suggests that $D^1$ no longer exists when the system approaches the density at which no or little shear is required to induce a finite pressure signifying that all particles are sterically hindered. Note that $D^1$ changes character already at packing fractions ($\sim 0.74$) far below the random close packing limit, which coincides with the lowest $\phi$ needed for shear induced jamming to occur \cite{ren13prl, Zhao2019_prl}. We can see this by probing the density dependent difference between $\mathit{MSD_{corr}}$ and $D^2_{min}$, as shown in Fig.~\ref{fig-phi}(d). For low $\phi$ we observe that these two local displacement metrics essentially coincide over the entire range of strains, whereas for larger $\phi$ we see a monotonic growth of their difference. This stems from the fact that the $D^1$ field gets more and more directionally heterogeneous at large $\gamma$ as the volume fraction increases towards random close packing.

To further test the uniqueness of the $D^1$ transport, we compute the vorticity $V$ of the non-affine grain flow (see S.M. \cite{sm}). We see that the value of $V^2$ averaged over the system, $\avg{V^2}$, first grows steadily with $\gamma$ and then reaches a plateau (see S.M. \cite{sm}). The strain at which $\avg{V^2}$ reaches the plateau, $\gamma_V$, shown in Fig.~\ref{fig-phi}(b) in comparison with $\gamma_A$, indicates that vortices develop and exist at a much longer time or strain scale than $D^1$ transport. This observation manifests the importance of $D^1$ transport: \emph{vorticity cannot capture} $D^1$. The plateau value of $\avg{V^2}$ first increases and then decreases as $\phi$ increases, (see Fig.~\ref{fig-phi}(c)). The change in trend occurs around $\sim 0.79$, which appears to be the point where $\gamma_A$ and $\gamma_V$ start to behave differently. Note that observing a trend change at this point seems to be consistent with results reported in works on glass transition in granular systems \cite{keys07natphys, candelier10epl} and the isotropic jamming point in frictional systems under compression \cite{silbert10sm, xiong19gm}.

\begin{figure}[t]
\centering
\includegraphics[width=\linewidth]{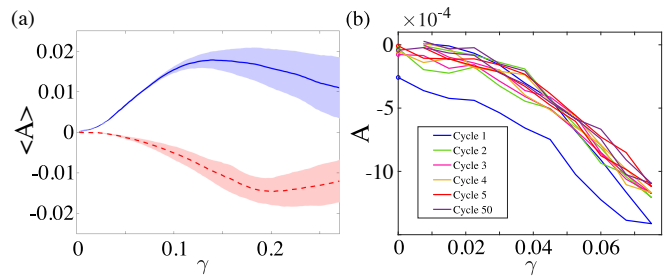}
\caption{(a) Examples of $D^1$ amplitude $A$ vs. shear strain $\gamma$ for two different initial shear cell shapes at $\phi = 0.773$, averaged over 5 runs: parallelogram (blue solid line) and rectangle (red dashed line). Shaded areas correspond to standard deviations. (b) $A$ vs. $\gamma$ for a packing undergoing cyclic shear ($\phi = 0.78$). Colors indicate different cycle number. Circles mark the end of corresponding cycles.}
\label{fig-cyclic}
\end{figure}

To highlight the subtle role of $D^1$ for amorphous packing dynamics, we probe how this transport mechanism depends on the initial conditions and responds to cyclic shear. As shown in Fig.~\ref{fig-cyclic}a, initiating shear from a rectangular box instead of a parallelogram \emph{inverts} the direction of $D^1$. This sensitivity to initial conditions makes $D^1$ a prime candidate for elucidating the emergence of memory in granular packings \cite{toiya04prl, fiocco14prl, royer15pnas, kou17nature}. While an in-depth study of $D^1$ in memory formation is left for future work, we can see in Fig.~\ref{fig-cyclic}a that the $D^1$ amplitude shows a distinct change in behavior beyond a critical strain amplitude, and note that memory formation in sheared amorphous packings is also strongly strain dependent. Another hallmark of structural memory formation is also observed in $D^1$ transport in Fig.~\ref{fig-cyclic}b. Transport over repeated cycles below the critical strain amplitude induces a relaxation towards limit cycle behavior. In previous work on cyclic shear dynamics in the same system, relaxation dynamics towards cyclic dynamics was also observed in the pressure of the packing. However, the pressure dynamics evolved towards a strain-symmetric response~\cite{ren13prl}. The absence of such relaxation dynamics in the microstructural dynamics as captured by $D^1$, and the sign change in its directionality depending on the initial conditions of the shear is striking. We see here a surprising disconnect between microstructural and pressure dynamics in amorphous packings and these observations amplify the importance of probing microstructural dynamics in amorphous packings, for example by looking at $D^1$.

\emph{Discussion ---} Our findings provide a new path to understanding the mechanisms of transport for sheared amorphous materials, especially at the transient stage where also shear jamming occurs~\cite{bi11nature,wang2018_prl}. Besides $D^2_{min}$, which we show to be greatly insensitive to packing conditions, there are clearly other transport mechanisms emerging that affect the dynamics of sheared amorphous particle packings. The disappearance of $D^1$ transport at larger $\phi$ yet below the random close packing limit indicates that there might be a third transport mechanism which cannot be well characterized within our experimental resolution. This clearly suggests new perspectives for numerical work: it would be interesting to see if the $D^1$ dynamics as observed in our model system can also be observed in numerical simulations; as they appear in the transient behavior, obtaining statistics should be computationally cheap comparing to those in the steady state. In fact, $D^1$ metrics can be used to efficiently verify the performance of numerical simulations, as they appear immediately after shear is initiated. Another question is how to capture $D^1$ and other such transport processes in models. Note that theoretical considerations~\cite{chacko18jfm,misra16cmt,poorsolhjouy19jmps} had already suggested that anisotropic displacement fields are a necessary ingredient to understand the dynamics of sheared amorphous system; our experimental data supports such perspectives.

\emph{Conclusions ---} We have studied particle transport in an amorphous particle packing and revealed that multiple distinct transport mechanisms occur at the same time. The experimental system in which we made these observations was composed of photoelastic disks exposed to quasi-static shear with various packing fractions $\phi$ and inter-particle friction coefficients $\mu$. We paid particular attention to the transient stage and found that there exist two distinct non-affine transport mechanisms in response to shear: ballistic random-walk-like diffusion that can be well characterized by $D^2_{min}$, and a second collective directional displacement phenomenon, which we denote as $D^1$ transport. The $D^2_{min}$ dynamics appears insensitive to either $\phi$ or $\mu$, while $D^1$ transport is sensitive to both. We find that when $D^1$ is subtracted from the displacement field, the trace of the diffusion tensor coincides with $D^2_{min}$, further solidifying the distinct nature of these two mechanisms. In the range of packing fractions just below random close packing, the $D^1$ correction to the overall displacement field does not isolate the diffusive motion very well anymore, suggesting that either the $D^1$ transport changes in character, or additional displacement mechanisms become relevant.

\begin{acknowledgments}

We are very grateful to discussions with Corey S. O'Hern, Brian P. Tighe, Lou Kondic, Joshua E. S. Socolar, and Pierre Ronceray. We greatly thank Robert P. Behringer for his initial support and intellectual input to this work. He unfortunately passed away before we could prepare any of this manuscript. This work was supported by NSFC Grant No. 41672256 (HZ), NSF Grants No. DMR1206351 and No. DMR1809762, and the W. M. Keck Foundation.

\end{acknowledgments}

\widetext
\clearpage
\begin{center}
\textbf{\large Supplemental Material: Sheared Amorphous Packings Display Two Separate Particle Transport Mechanisms}
\end{center}
\setcounter{equation}{0}
\setcounter{figure}{0}
\setcounter{table}{0}
\setcounter{page}{1}
\makeatletter
\renewcommand{\theequation}{S\arabic{equation}}
\renewcommand{\thefigure}{S\arabic{figure}}
\renewcommand{\bibnumfmt}[1]{[S#1]}
\renewcommand{\citenumfont}[1]{S#1}

This supplemental material provides: \rm{I} a detailed description of the experimental setup and examples showing photoelastic particles with different friction coefficients; \rm{II} results of $\mathit{MSD_{corr}}$ after correcting non-affine displacements by only subtracting the $\avg{\Delta x}(x)$ term but not the $\avg{\Delta y}(x)$ term; \rm{III} diffusion anisotropy determined from the $\avg{\Delta x}_y$- and $\avg{\Delta y}_y$-corrected diffusion tensor $T$; \rm{IV} results showing the ballistic behavior of system-averaged $D^2_{min}$ as a function of the deviatoric shear strain that is calculated from $D^2_{min}$ fitting; \rm{V} Details about determining the $D^1$ transport amplitude $A$; \rm{VI} detailed description for vorticity calculation and the associated behavior of vorticity; and \rm{VII} comparison between diffusion amplitudes extracted from $\mathit{MSD_{corr}}$ and $D^2_{min}$ at various $\phi$ and $\mu$.

\section{Experimental Setup}

\begin{figure}[b]
\centering
\includegraphics[width=0.65\linewidth]{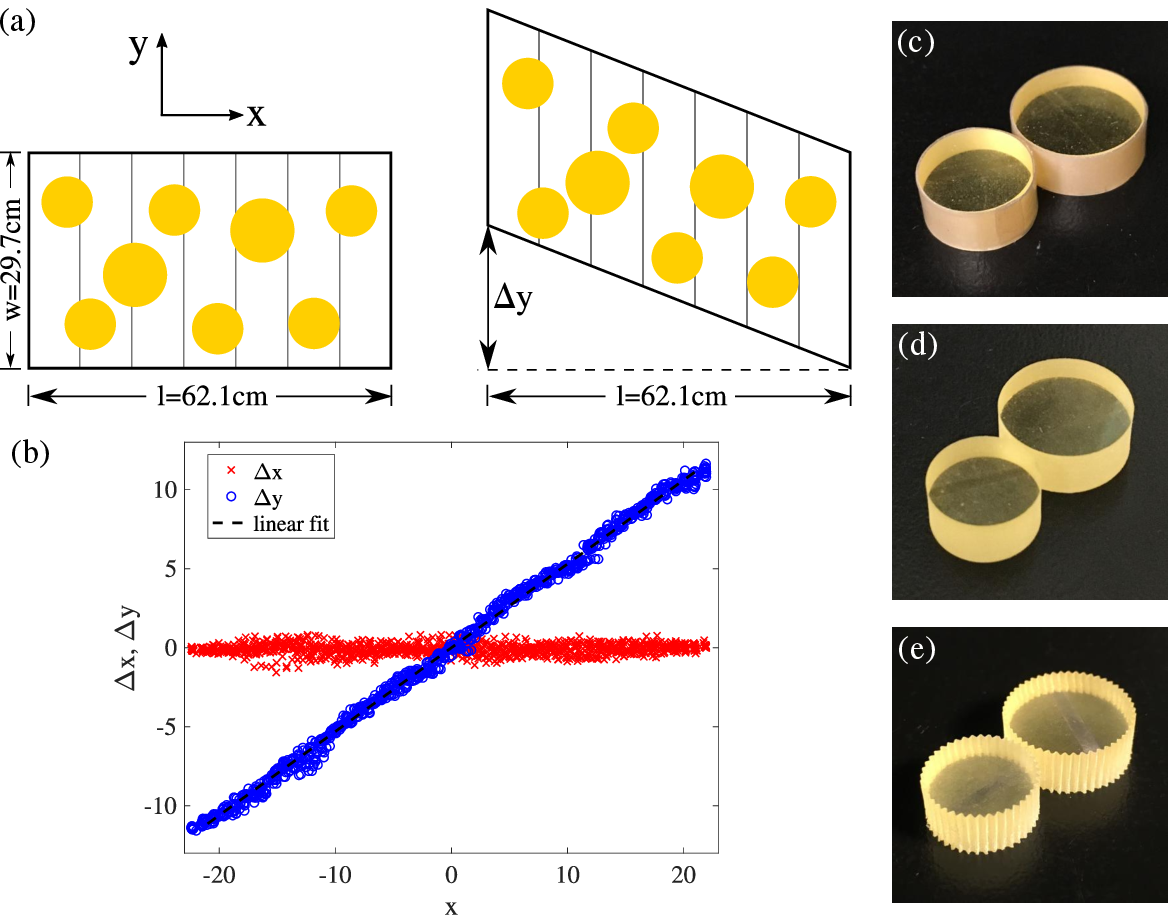}
\caption{(a) Schematic of special apparatus for applying simple shear strain to a collection of particles, here photoelastic discs. The system base consists of narrow slats that are the same size as the smaller particle diameter. When the sidewalls deform to create simple shear, the slats move with the walls so that the base deforms affinely. In the sketch, both the slats and particles are drawn much larger relative to the boundaries than in the real experiment. The $x-y$ axes indicate the coordinate system in the lab frame, where simple shear is applied along the y axis, and the shear strain, $\gamma$, is defined as $\gamma = \Delta y/l$. (b) Examples of total particle displacements in the $x$ ($\Delta x$, crosses) and $y$ ($\Delta y$, circles) direction as a function of particle $x$ position for a packing at packing fraction 0.8 after a shear strain of 0.054. The black dashed line shows a linear fit of $\Delta y$ as a function of $x$. All lengths are in unit of the average particle diameter. (c - e) Close-up examples showing the particle side property with different friction coefficients, from low to high.}
\label{fig-apparatus}
\end{figure}

The experiment is designed to provide a homogeneous linear shear profile without shear bands, as discussed in Ren \textit{et al.} \cite{ren13prl}. A sketch of this setup is shown in Fig.~\ref{fig-apparatus}(a). The key aspect of this apparatus is that the base, which consists of a set of narrow slats, deforms affinely with the confining boundaries. The whole is driven by a linear motor, and the result corresponds to deforming a rectangle into a parallelogram of the same area. Each of the narrow slats that form the base has a width equal to the smallest particle diameter. The slats are coated with a fine powder to reduce friction with the particles, but there is weak residual friction. Consequently, rattlers are carried with the applied strain, unlike what happens in a conventional shear cell, where rattlers remain fixed until they collide with other particles. In very soft systems near jamming for a conventional experiment, rattlers that remain at rest tend to lead to the creation of important density inhomogeneities that may be system spanning. This does not occur for the present experiment, as shown by Ren \textit{et al.} \cite{ren13prl}. As an example, Fig.~\ref{fig-apparatus}(b) shows particle displacements in both $x$ ($\Delta x$) and $y$ ($\Delta y$) direction as a function of particle $x$ position after a shear strain $\gamma = 0.054$ is applied. We can clearly see the majority of particle displacements occur in the $y$ direction and $\Delta y$ behaves almost linearly with respect to $x$ position. Note that Fig.~1(c) in the main text is reproduced by bin averaging the $\Delta x$ data.

Fig.~\ref{fig-apparatus}(c-e) show examples of photoelastic particles with different friction coefficients ($\mu$) used in the experiments: (b) particles with Teflon tape wrapped around edges, resulting with $\mu \approx 0.15$; (c) particles normally cut from flat photoelastic sheets with $\mu \approx 0.7$; (c) particles cut by water jet to have fine teeth on the edge, providing an extremely high $\mu$. We refer to these three friction coefficients as $\mu_l, \mu_m$, and $\mu_h$, respectively.


\section{MSD corrected only by $\mathbf{\avg{\Delta x}_y(x)}$}

\begin{figure}
\centering
\includegraphics[width=0.75\linewidth]{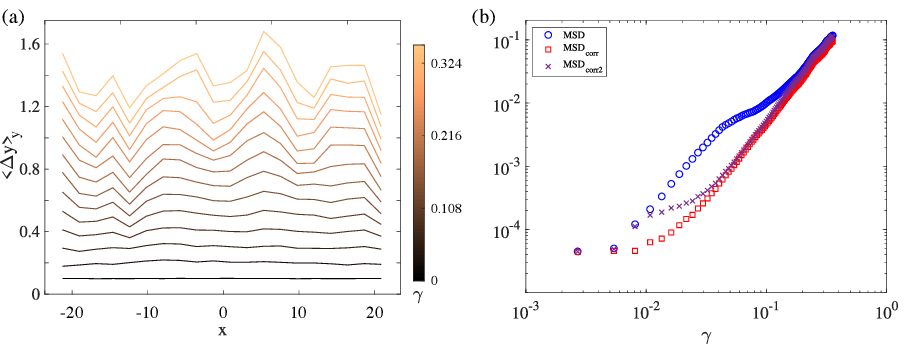}
\caption{(a) Non-affine $y$ displacements $\Delta y$, averaged in the same way as Fig.~1(c) in the main text, as a function of the bin center at different shear strains. From bottom up, a shear strain $\gamma$ of $0.027$ is applied between two consecutive lines, where the first line is at $\gamma = 0.0027$, and each upper line is added by $0.1$ for better presentation. (b) Diffusion matrix tensor trace $MSD$ before (circles, denoted as $\mathit{MSD}$), after subtracting both $\avg{\Delta x}_y$ and $\avg{\Delta y}_y$ (squares, denoted as $\mathit{MSD_{corr}}$), and after subtracting only $\avg{\Delta x}_y$ but not $\avg{\Delta y}_y$ from the non-affine displacements (crosses, denoted as $\mathit{MSD_{corr2}}$), vs. $\gamma$. Data from the same run in Fig.~1 in the main text. All lengths are in unit of the average particle diameter.}
\label{fig-msd}
\end{figure}

To show that the importance of large scale collective transport is more pronounced in the $x$ direction than in the $y$ direction, we here calculate the trace of the diffusion tensor $T$ (see main text Eq.~(1)). We computed the diffusion tensor by three methods: (1) without any correction to the non-affine displacements, (2) with corrections to both $x$ and $y$ non-affine displacements, by subtracting $\avg{\Delta x}_y(x)$ and $\avg{\Delta y}_y(x)$, respectively; and (3) with corrections to only $x$ non-affine displacements by subtracting $\avg{\Delta x}_y(x)$. We refer to the trace calculated from these three methods as $\mathit{MSD, MSD_{corr}}$, and $\mathit{MSD_{corr2}}$, respectively. Fig.~\ref{fig-msd}(a) shows the large scale transport in $y$ direction, as a comparison to Fig.~2(a) in the main text. Note that the amplitude of this kind of transport in $y$ direction is much smaller than that in $x$ direction. Fig.~\ref{fig-msd}(b) compares results of $\mathit{MSD, MSD_{corr}}$, and $\mathit{MSD_{corr2}}$, and we can see that $\mathit{MSD}$ clearly differs from $\mathit{MSD_{corr}}$ and $\mathit{MSD_{corr2}}$, while $\mathit{MSD_{corr}}$ and $\mathit{MSD_{corr2}}$ are very close to each other. Therefore, we conclude that the effect of $\avg{\Delta y}_y(x)$ is negligible.


\begin{figure}
\centering
\includegraphics[width=0.8\linewidth]{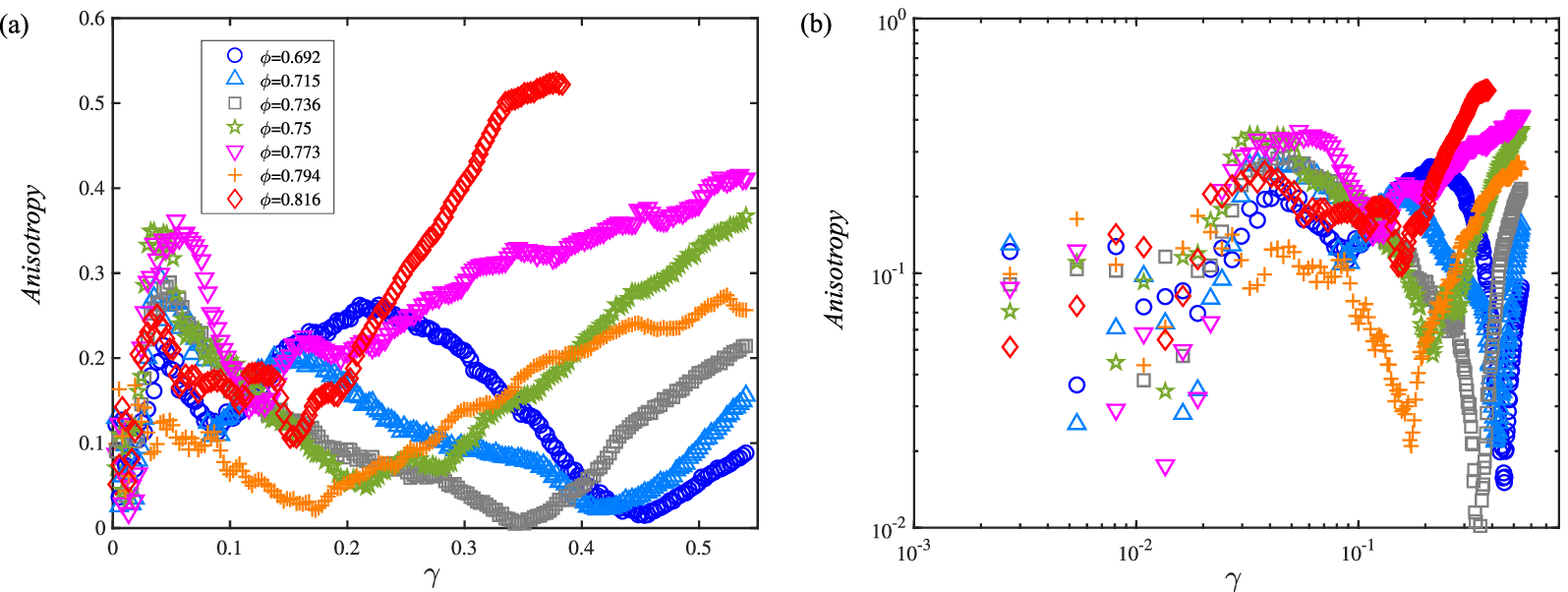}
\caption{Diffusion anisotropy in the $\mu_m$ particle system, plotted as a function of shear strain $\gamma$ in a (a) linear-linear scale, and (b) log-log scale. Different colors and symbols indicate different $\phi$. For each $\phi$, one typical run was selected to be plotted.}
\label{fig-anis}
\end{figure}

\section{Diffusion anisotropy}
We show in Fig.~\ref{fig-anis} examples of the evolution of the diffusion anisotropy as a function of $\gamma$ for various $\phi$. The diffusion anisotropy is defined as the absolute value of the difference between the eigenvalues of the diffusion tensor $T$, normalized by the trace of $T$. We can see from Fig.~\ref{fig-anis} that there still exists finite anisotropy even after we remove the $D^1$ transport from the non-affine displacements. Especially at larger strain, the diffusion anisotropy seem to keep increasing after a certain $\gamma$, regardless of $\phi$. We attribute this to the insufficiency of merely subtracting $D^1$ transport from the non-affine motion, which is qualitatively consistent with our observation of the behavior of $\mathit{MSD_{corr}}$.


\begin{figure}
\centering
\includegraphics[width=0.5\linewidth]{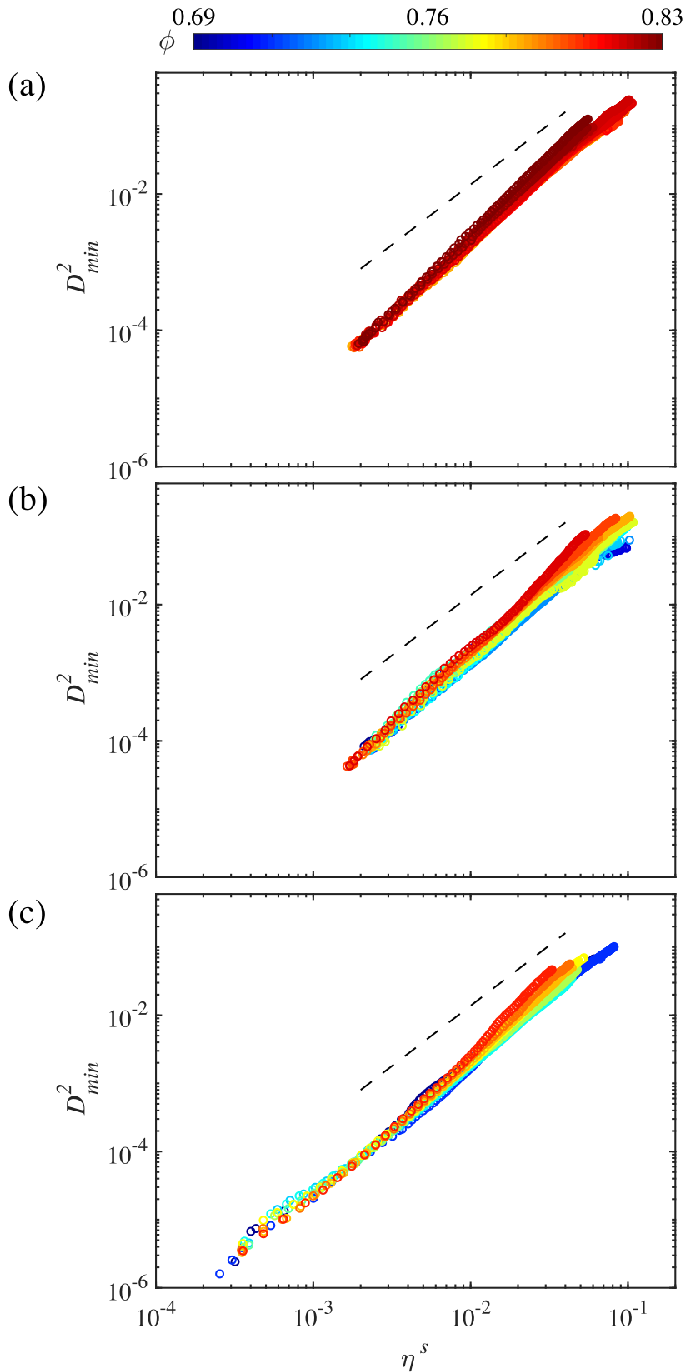}
\caption{$D^2_{min}$ vs. the deviatoric shear strain $\eta^s$ obtained from the best fit affine strain tensor $\mathbf{J}_i$, both averaged over particles at a given applied bulk shear strain for (a) $\mu_l$, (b) $\mu_m$, and (c) $\mu_h$, particle systems. Different $\phi$ are indicated by the same color bar for all $\mu$. The dashed black lines have a slope of 2 and are to guide the eye. All lengths are in unit of the average particle diameter.}
\label{fig-eta}
\end{figure}

\begin{figure}[b]
\centering
\includegraphics[width=0.95\linewidth]{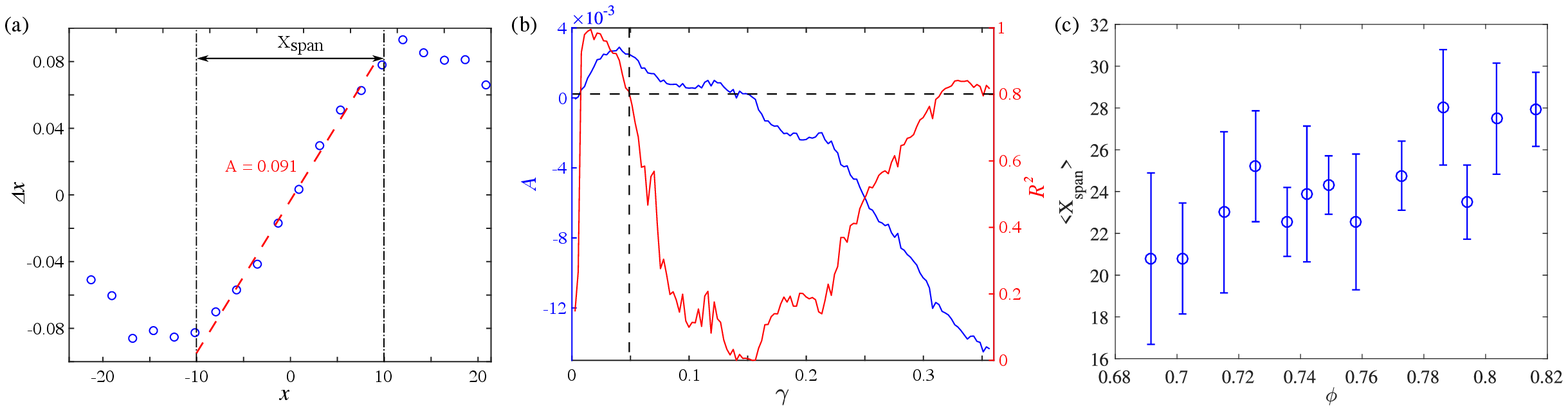}
\caption{(a) An example to show how $D^1$ amplitude $A$ is obtained from the relationship between bin-averaged non-affine $x$ displacement $\avg{\Delta x}_y$ and the bin center position $x$. The black dash-dotted lines indicate the bounds between which the linear fit is performed with the size of the bound denoted as $X_{span}$, and the red line shows the linear fit result with the $A$ value. Data is the same as in Fig.~1(c) in the main text. (b) Left: $D^1$ transport amplitude $A$ vs. $\gamma$. Right: corresponding R-squared value $R^2$ using the fitting method described in (a) vs. $\gamma$. The horizontal dashed line corresponds to $R^2 = 0.8$. (c) Average $\avg{X_{span}}$ over 5 runs as a function of packing fraction $\phi$ for $\mu_m$ particles. All lengths are in unit of the average particle diameter.}
\label{fig-slope}
\end{figure}

\section{$\mathbf{D^2_{min}}$ and deviatoric shear strain}

Since being proposed by Falk and Langer \cite{falk98pre}, $D^2_{min}$ measurements have been applied widely to study irreversible deformation and avalanches in amorphous systems, metallic glasses, and so on. It has been shown that larger non-affine deformation correspond to larger local $D^2_{min}$. Statistically, Li \textit{et. al} \cite{li15pre} have shown that $D^2_{min}$, when averaged over the system, exhibits random walk like behavior as a function of the deviatoric shear strain. Here we see similar results in the quasi-statically sheared granular systems at various packing fractions and particle friction coefficients.
As shown in Eq.~1 in the main text, an affine strain tensor $\mathbf{J}_i$ is fitted to minimize $D^2$ value for particle $i$ as an estimate for the background deformation field, which is then used to calculate $D^2_{min}$. From this best fit local strain field, the deviatoric shear strain $\eta^s(\gamma)$ at a given bulk shear strain $\gamma$ can be calculated in the following way:
\begin{equation}
    \eta^s(\gamma) = \avg{\eta_i^s(\gamma)} = \avg{ \frac{1}{2}\sqrt{Tr((\bm{J}_i^T\bm{J}_i)^2)-Tr(\bm{J}_i^T\bm{J}_i)^2}},
\end{equation}
where $\mathbf{J}_i^T$ is the transpose of $\mathbf{J}_i$. As shown in Fig.~\ref{fig-eta}, $D^2_{min}$ show a ballistic behavior as a function of $\eta^s$, and this seems to be universal regardless of packing fractions $\phi$ and inter-particle friction coefficients $\mu$. In a random walk model, the mean squared displacement displays ballistic behavior at smaller time scales and diffusive behavior at larger time scales. We attribute the observation of seeing only ballistic behavior in $D^2_{min}$ to $\eta^s$ being small.

\section{$\mathbf{D^1}$ Transport Amplitude}

We show here a detailed description for how we extract $D^1$ transport amplitude $A$ shown in the main text. Fig.~\ref{fig-slope}(a) shows the same set of data in Fig.~1(c) in the main text without the error bar. To obtain $A$, we linearly fit $\avg{\Delta x}_y$ with $x$ in the range $x$ indicated by the two black dash-dotted lines in Fig.~\ref{fig-slope}(a), and the slope reveals $A$. The size of the $x$ range over which the linear fit is performed is denoted as $X_{span}$.

It is worth noting that the linear fit is a first order approximation to describe the behavior of $D^1$ transport as a function of $x$. The goodness of the fit, indicated by the standard R-squared value $R^2$, is shown in Fig.~\ref{fig-slope}(b), along with the amplitude $A$, as a function of shear strain $\gamma$. We can see that the linear fit serves as a good description for small $\gamma$ and fails to do so at large $\gamma$ until the end stage of shear. This failure presumably is related to the complex nature of particle transport and the linear nature of $D^1$ transport as a function of $x$ no longer holds. In Fig.~3, we only show data at small $\gamma$ where $R^2$ is greater than $0.8$.

In addition, we measure $X_{span}$ when $A = A_{max}$ for all the runs and we show the results in Fig.~\ref{fig-slope}(c), averaged over 5 runs for each $\phi$. Fig.~\ref{fig-slope}(c) shows that $\avg{X_{span}}$ is in the range between $20$ and $30$ particles, much larger than the size of a typical shear transformation zone, which spans over just a few particles. In addition, $\avg{X_{span}}$ shows little dependence on $\phi$. Note that there are approximately $45$ particles spanning across the $x$ direction in our experiments. Hence we believe the effect on $D^1$ transport from the wall is minimal, which usually extends $3 - 5$ particles \cite{suzuki08apt, dupont03epl}.


\section{Vorticity}

\begin{figure}
\centering
\includegraphics[width=0.95\linewidth]{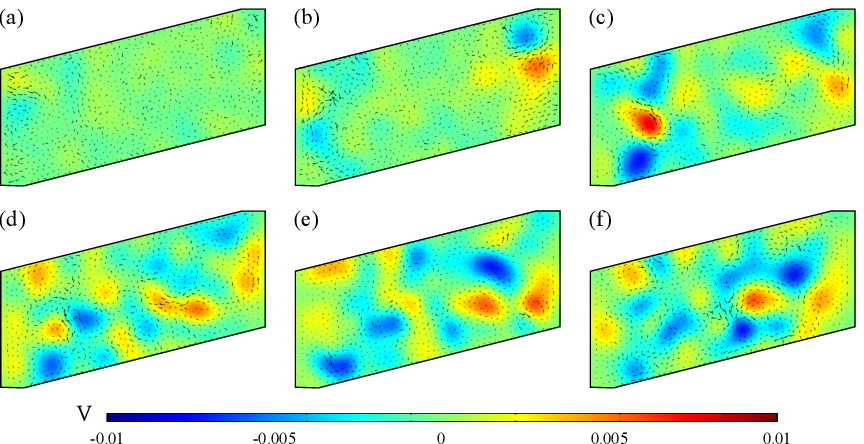}
\caption{Coarse grained vorticity in the system with $\phi = 0.796$ at different shear strain $\gamma$: (a) $\gamma = 0.0081$, (b) $\gamma = 0.0378$, (c) $\gamma = 0.0675$, (d) $\gamma = 0.0972$, (e) $\gamma = 0.1269$, and (f) $\gamma = 0.1566$. The black arrows show particle non-affine velocity (rescaled for visual guidance). All lengths are in unit of the average particle diameter.}
\label{fig-vorticity_example}
\end{figure}

We include here a detailed description calculations related to vorticity $\mathbf{V}$. To calculate $\mathbf{V}$, we first calculate non-affine particle velocities at a given strain by dividing their non-affine displacements from $3$ strain steps backward to $3$ strain steps forward by $6$ (we set one shear strain step to be the unit time). We then mesh the shear box with the mesh grid size to be approximately the small particle radius. The velocity field $\mathbf{v}$ at these mesh grids can be interpolated by cubic interpolation. Finally the vorticity $\mathbf{V}$ simply follows the definition $\mathbf{V}(\gamma) = \nabla \times \mathbf{v}$. Figure~\ref{fig-vorticity_example} coarse grained $\mathbf{V}$ at different $\gamma$ at $\phi = 0.79$ as an example.

\begin{figure}
\centering
\includegraphics[width=0.95\linewidth]{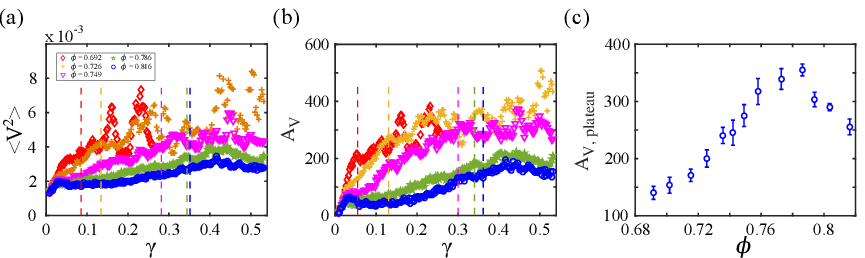}
\caption{(a) Examples of the average vorticity squared in the system, $\avg{V^2}$, as a function of shear strain $\gamma$. Different symbols and colors indicate different packing fraction $\phi$. The dashed lines correspond to the determined strain for $\avg{V^2}$ to reach the plateau. (b) Examples of the area with vorticity greater than the selected threshold, $A_V$, as a function of $\gamma$. $\phi$ is indicated in the same way as (a). The dashed lines correspond to the determined strain for $\avg{V^2}$ to reach the plateau. (c) Plateau value of $A_V$ normalized by the average diameter squared, $A_{V, plateau}$, as a function of $\phi$. Each data point results from averaging over 5 runs for each $\phi$. All lengths are in unit of the average particle diameter.}
\label{fig-vorticity}
\end{figure}

The amplitude of $\mathbf{V}$ generally first increases with shear strain $
\gamma$ and then reaches a plateau, as shown in Fig.~\ref{fig-vorticity}(a) for the average value of $\mathbf{V}^2$. We characterize this behavior by measuring the strain needed for $\mathbf{V}^2$ to reach the plateau ($\gamma_V$) and the plateau value (defined as the average value of $\mathbf{V}^2$ at the plateau stage), measured at each run and averaged over $5$ runs at any given $\phi$. These results have been shown in the main text as Figs.~3(b-c). Note that at low $\phi$ (Fig.~\ref{fig-vorticity}), $\mathbf{V}^2$ seems to increase fast at small $\gamma$ and then keep increasing at a much lower speed before it reaches the plateau. $\gamma_V$ is measured after the second phase of increase in this case.

We further verify the dependence of $\gamma_V$ and $\mathbf{V}^2_{plateau}$ on $\phi$ by measuring the area in the system with $|\mathbf{V}|$ greater than a certain threshold. Here we choose the threshold to be the average of $|\mathbf{V}|$ from all the strain steps at the lowest $\phi$. This area value, denoted as $A_V$, shows very similar behavior to that of $\mathbf{V}^2$, as shown in Fig.~\ref{fig-vorticity}(b). In Fig.~\ref{fig-vorticity}(c), we show the plateau value of $A_V$, $A_{V, plateau}$ as a function of $\phi$. Similar to that of $\mathbf{V}^2_{plateau}$, this area plateau value first increases and then decreases as $\phi$ increases, and the transition seems to happen around $0.79$, consistent with all the other measurements we have shown in the main text.

\begin{figure}
\centering
\includegraphics[width=0.75\linewidth]{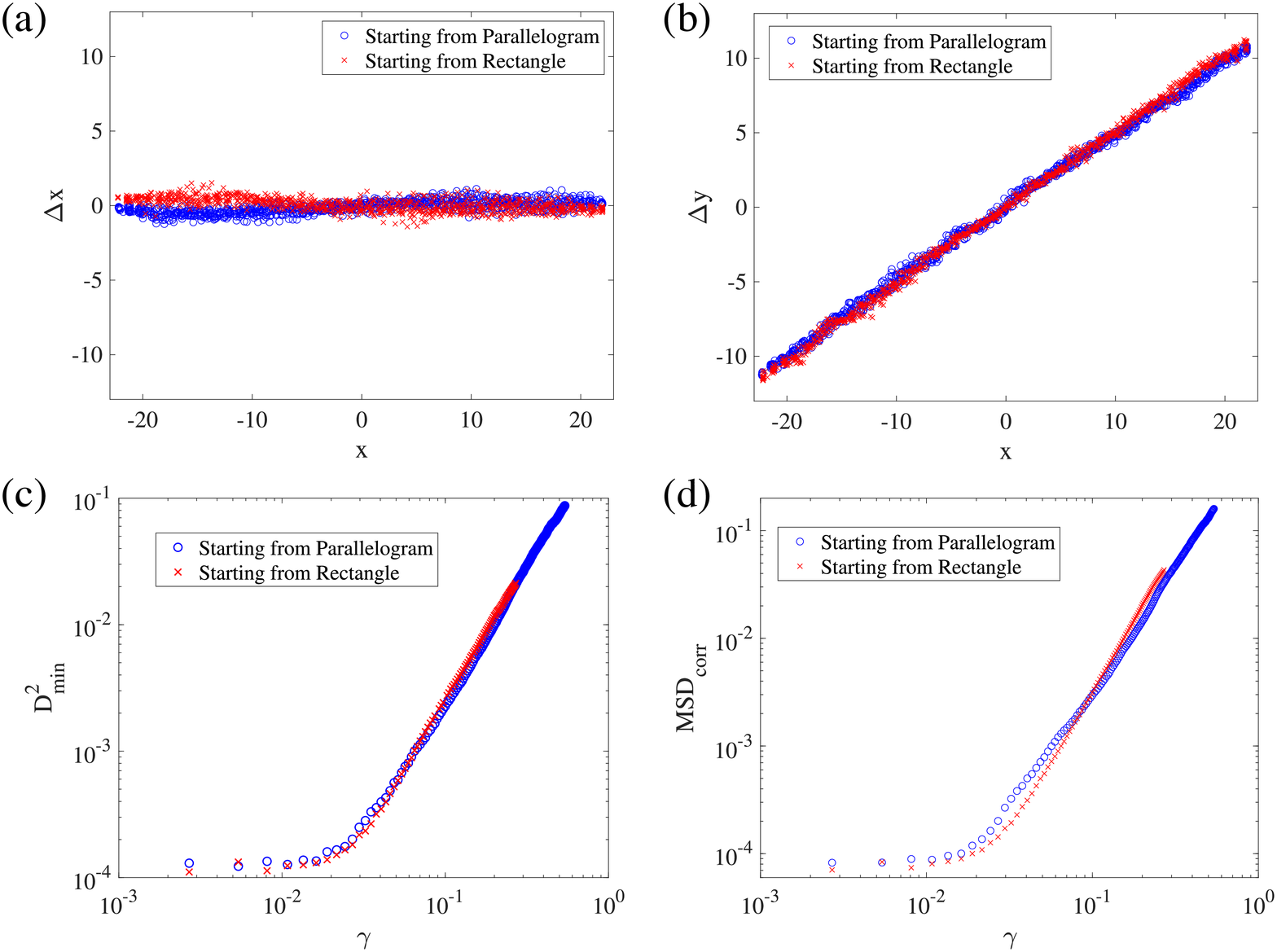}
\caption{Total displacements of particles in: (a) $x$ ($\Delta x$), and (b) $y$ ($\Delta y$) directions, as a function of particle $x$ position. (c) $D^2_{min}$, and panel (d) shows $\mathit{MSD}_{corr}$ after taking out the linearly fitted shear profile, $\avg{\Delta x}_x$, and $\avg{\Delta y}_x$. In (a - d), blue circles are data from an experiment where the starting configuration of the shear box is a parallelogram and red crosses are data from an experiment where the starting configuration of the shear box is a rectangle. Both experiments have a packing fraction of 0.78 and the data are obtained at shear strain $\gamma = 0.1485$.}
\label{fig-bc}
\end{figure}


\section{Effect of Boundary Conditions}

We show here in Fig.~\ref{fig-bc} a direct comparison between two boundary conditions presented in Fig.~4(a) in the main text. Two runs at $\phi = 0.78$ and $\gamma = 0.1485$ are shown, one starting from a parallelogram shear box and one starting from a rectangle shear box. Fig.~\ref{fig-bc}(a) shows particle displacements in the $x$ direction, which are used to bin-average to obtain $D^1$ dynamics and $D^1$ slope $A$. As shown in Fig.~4(a) in the main text, the amplitude of $A$ is comparable for two boundary conditions, while the sign of $A$ is opposite. Fig.~\ref{fig-bc}(b) shows particle displacements in the $y$ direction as a function of particle position $x$, exhibiting a fine linear relationship with $x$. Hence the suppression of shear bands works in both boundary conditions. Fig.~\ref{fig-bc}(c) and (d) show the comparison of diffusion behaviors. While $D^2_{min}$ seems almost unaffected by changing the boundary condition (Fig.~\ref{fig-bc}(c)), we see a slight difference in $\mathit{MSD}_{corr}$ for two boundary conditions after removing linear shear profile fitting and correcting $\avg{\Delta x}_y(x)$ and $\avg{\Delta y}_y(x)$. We postulate that the latter discrepancy is caused by different behaviors exhibited by $D^1$ transport under different boundary conditions. Therefore, we conclude that by changing the starting shear box boundary condition, only $D^1$ transport mechanism is affected while the linear simple shear and the diffusion transport mechanisms are not affected.

\section{Diffusion Exponent and Amplitudes}

\begin{figure}[b]
\centering
\includegraphics[width=0.6\linewidth]{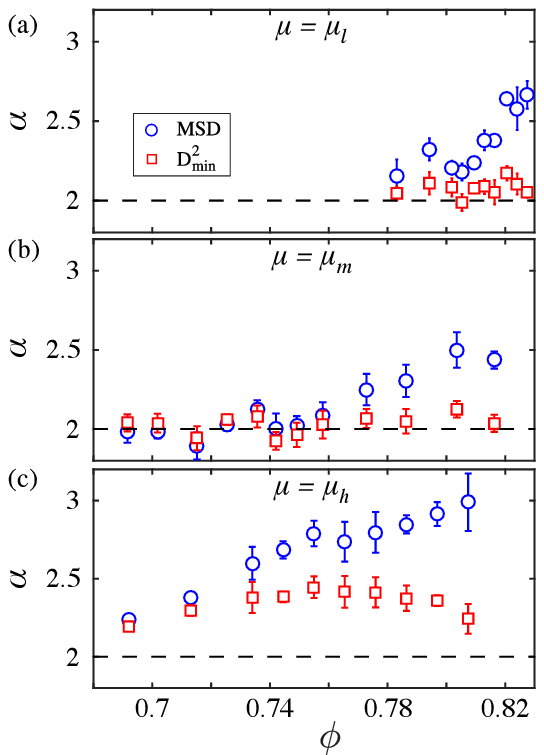}
\caption{Diffusion exponent $\alpha$ as a function of the system packing fraction for low (a), medium (b) and high (c) intergranular friction. Squared red symbol id for measure with $D^2_{\rm{min}}$ and blue circle symbols is for measure with MSD. Vertical bars show the $95\%$ confident interval for the slope measurement.}
\label{fig-mu}
\end{figure}

We can quantify the difference of $\mathit{MSD_{corr}}$ and $D^2_{min}$ by quantifying their strain dependence. In particular, we measure the exponent $\alpha$ and the amplitude $D_0$ from both $\mathit{MSD_{corr}}$ and $D^2_{min}$:
\begin{equation}
    D^2_{min}, \mathit{MSD_{corr}} = D_0 \gamma^\alpha.
    \label{eq-msd}
\end{equation}
For the $\mu_m$ particles, we show the results in Fig.~\ref{fig-mu}(b). We observe that the exponent for $D^2_{min}$ is independent of $\phi$ and indistinguishable from 2 ~\cite{li15pre}, which in our interpretation means that the background field-insensitive diffusive displacements is a permanent feature of these disordered materials. When we look at $\mathit{MSD_{corr}}$, we see that this exponent starts to peel away from 2 at higher volume fractions, implying again that the mesoscopic background displacement correction we apply is not anymore sufficient to remove all background displacement and make the diffusion tensor equal in amplitude to the $D^2_{min}$. Note that the amplitude prefactor of the power law behavior in displacements gives a similar indication. Assuming that no additional transport mechanisms emerge, the $\mathit{MSD_{corr}}$ behavior suggests that $D^1$ is changing character just before disappearing entirely at random close packing.

\begin{figure}
\centering
\includegraphics[width=0.45\linewidth]{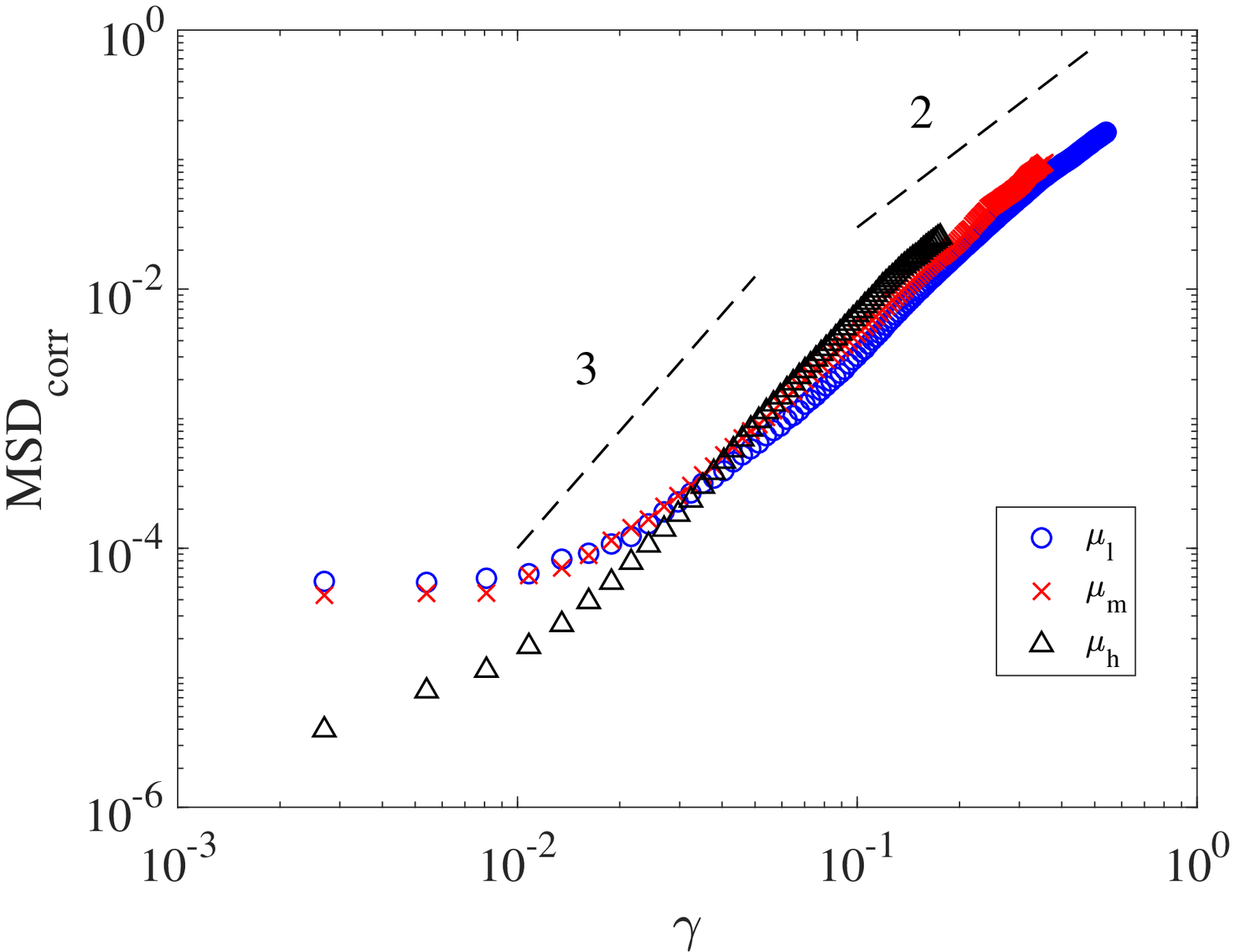}
\caption{Examples of $\mathit{MSD_{corr}}$ for three friction particles. The packing fraction for each type particle is $0.794, 0.794$, and $0.797$, for $\mu$ from low to high. Black dashed lines are for eye guidance with denoted slopes. All lengths are in unit of the average particle diameter.}
\label{fig-msdcomp}
\end{figure}

An additional observation is that the diffusion exponent is much larger for the $\mu_h$ particles. To understand this, we plot $\mathit{MSD_{corr}}$ for three $\mu$ particles at roughly the same $\phi \approx 0.79$, as shown in Fig.~\ref{fig-msdcomp}. We speculate that the much higher $\alpha$ value for $\mu_h$ is a small shear strain effect, i.e., $\gamma$ at which $\alpha$ is measured is rather small. Note that in Fig.~\ref{fig-msdcomp}, $\mathit{MSD_{corr}}$ for $\mu_h$ particle appears to show an exponent much closer to $2$ at larger shear strains. However, due to experiment limit, we do not have additional data for larger $\gamma$ to support such a speculation.

\begin{figure}
\centering
\includegraphics[width=0.47\linewidth]{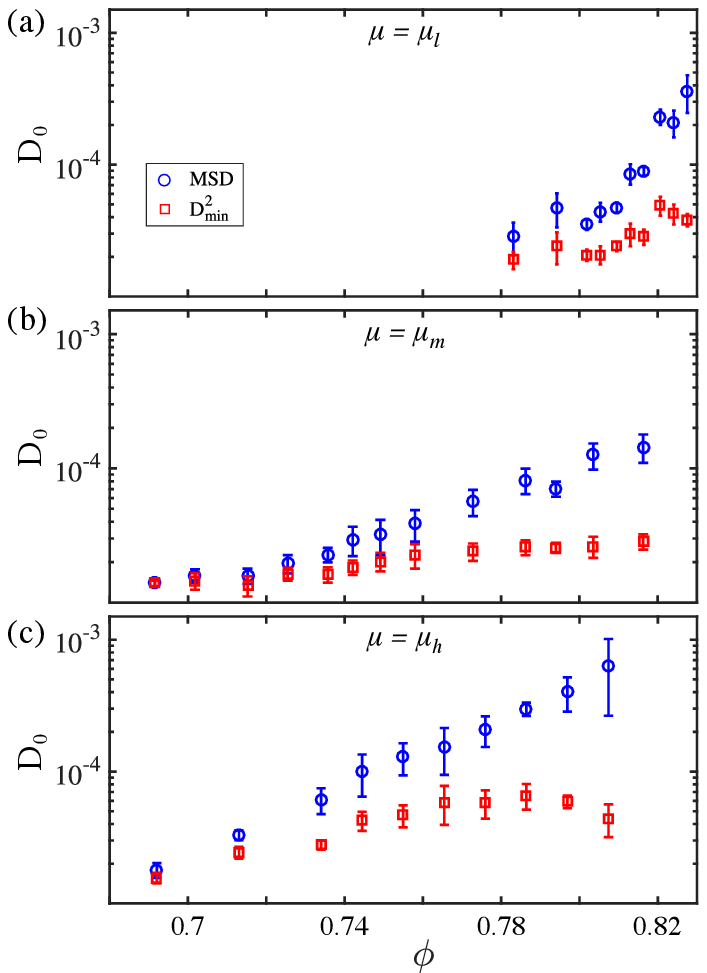}
\caption{Diffusion amplitude $D_{0}$ extracted from $\mathit{MSD_{corr}}$ (circles) and $D^2_{min}$ (squares) vs. packing fraction $\phi$, with each $\phi$ averaged over $5$ runs, for (a) $\mu_l$, (b) $\mu_m$, and (c) $\mu_h$, particle systems. All lengths are in unit of the average particle diameter.}
\label{fig-amplitude}
\end{figure}

It is known that the friction coefficient affects the window of $\phi$ where nontrivial packing mechanics occurs \cite{silbert10sm,luding2016nature,xiong19gm}. For lower $\mu$, the window is expected to start at higher $\phi$. We may expect that $D^1$ dynamics is similarly affected. If we lower $\mu$ of the particles by wrapping them with Teflon tape, the onset of shear jamming moves to higher $\phi$. Similarly, we expect the difference of $\mathit{MSD_{corr}}$ and $D^2_{min}$ to emerge at higher densities. This is indeed what we observe: we extract the $\mathit{MSD_{corr}}$ and $D^2_{min}$ exponent for low friction particles and observe indeed that the $\mathit{MSD_{corr}}$ starts to deviate from $D^2_{min}$ at higher $\phi$ as compared to the experiments done at $\mu_m$: see Fig.~\ref{fig-mu}(a). If we then perform the same shear experiments with gear-shaped particles with essentially \textit{infinite} friction coefficient, the range of $\phi$ over which $\mathit{MSD_{corr}}$ is different from $D^2_{min}$ has expanded to lower $\phi$ as shown in Fig.~2(c) in the main manuscript. Note that for the $\mu_h$ particles, the diffusive exponents are slightly higher than two and have become weakly sensitive to the packing fraction.

As we fit $D^2_{min}$ and $\mathit{MSD_{corr}}$ as a function of shear strain $\gamma$ via Eq.~\ref{eq-msd}, we can also probe the behavior of $D_0(\phi,\mu)$. The measurement of $\alpha$ for $\mathit{MSD_{corr}}$ starts to deviate from that for $D^2_{min}$ as $\phi$ increases, as shown in Fig.~\ref{fig-mu}. We interpret this deviation as an indication of the inadequacy for merely using $D^1$ transport to capture the large scale non-affine particle motion. Here in Fig.~\ref{fig-amplitude}, similar dependence on $\phi$ can also be observed for $D_0$ for all friction coefficients tested in our experiments.

\end{document}